# On the harmonic superspace language adapted to the Gelfand-Dickey algebra of differential operators

**M. HSSAINI, M. KESSABI, B. MAROUFI**
UFR-HEP, Section Physique des Hautes Energies, Département de Physique
Faculté des sciences, B.P.1014, Rabat, Morocco*
and
**M.B.SEDRA**[*,+]
Laboratoire de Physique Théorique et Appliquée (L.P.T.A.) Département de
Physique, Faculté des sciences, B. P.133, Kénitra, Morocco

**Abstract**

Methods developed for the analysis of non-linear integrable models are used in the harmonic superspace (HS) framework. These methods, when applied to the HS, can lead to extract more information about the meaning of integrability in non-linear physical problems. Among the results obtained, we give the basic ingredients towards building in the HS language the analogue of the G.D. algebra of pseudo-differential operators. Some useful convention notations and algebraic structures are also introduced to make the use of the harmonic superspace techniques more accessible.

# 1. Introduction

As its well known, conformal symmetry and its higher spin extensions have played a central role in the study of string dynamics, statistical models of critical phenomena and more generally in two dimensional conformal field theories (CFT) [1]. These are symmetries generated by conformal spin s currents with $s \leq 2$. Some years ago, a new issue of possible extensions of the conformal symmetry to higher conformal spins was opened by Zamolodchikov [2]. Among his results, was the discovery of a new algebra, $W_3$, involving besides the usual spin-2 energy momentum tensor, a conformal spin-3 conserved current. This symmetry, which initially was identified as the symmetry of the critical three states Potts model, has also been realised as the gauge symmetry of the so-called $W_3$-gravity[3]. Since then higher spin extensions of the conformal symmetry have been studied intensively by using different methods: field theoretical, Lie algebraic, geometrical approach or again by using infinite matrices. $W_3$-algebras and their supersymmetric extensions are also known to arise from the second Hamiltonian structure of the generalised Korteweig-de Vries (KdV) hierarchy [4]. A common feature of the higher spin extended algebras is that they can be used to describe the underlying symmetries of non-linear integrable hierarchies [5].

Methods developed for the analysis of integrable models can be also used to study various important physical problems. Focussing for the moment one of the important field of applications, namely the hyper-Kahler metrics building program. This is an important question of hyper-Kahler geometry (which can be solved in a nice way in the harmonic superspace [6,7]) a subject much studied in modern theoretical physics, more especially in connection with the theory of gravitational instantons; moduli problems in monopole physics, string theory and elsewhere [8,9,10].

Following a program already started in [11,12,13] and which consist in using the H.S. technology to study various physical problems, we propose in the present work to build the "huge algebra" of pseudo-differential operators defined on the sphere $S^2 = SU(2)/U(1)$. Such operators will be called for simplicity harmonic operators and are suspected to give rise to extended Lax operators.

To that purpose we shall follow the following steps. Introducing in (section 2) the ring of "analytic" harmonic functions, we focus in (section 3) to present the basic steps towards building, in the harmonic superspace language, the analogue of the higher conformal spin algebra of pseudo-differential operators [14]. We show in particular that any harmonic pseudo-differential operator is completely specified by an "harmonic weight" $\Delta$, two integers p and q = p + n, n>0, defining respectively the lowest and highest degrees and finally a finite number of analytic harmonic fields belonging to the ring R.

We introduce the space of differential operators of fixed harmonic weight and fixed degrees. This space is refereed to as $\Xi_\Delta^{(p,q)}$. Then we consider the set of non-linear operators of fixed degrees but arbitrary harmonic weight. Finally we build the desired space and present some points to discussion.

## 2. The Ring of harmonic functions

Let us first consider the ring of functions, defined on the sphere $S^2 = SU(2)/U(1)$, which are assumed to expand in series of the symmetrised product of harmonic variables $(u^+, u^-)$ as

$$F_{[n+,m-]}(u) = f^{(i_1 \ldots i_n j_1 \ldots j_m)} u^+_{(i_1} \ldots u^+_{i_n} u^-_{j_1} \ldots u^-_{j_m)}, \tag{2.1}$$

where we have introduced the convention notation [n+, m-] to describe respectively the number of harmonic variables $u^+$ and $u^-$ in the symmetrised harmonic expansion Eq.(2.1) for arbitrary integers n and m. This equation yields among others to introduce an "harmonic weight" denoted by $\Delta$ and which we define as

$$\Delta F_{[n+,m-]}(u) = (n,m), \tag{2.2}$$

and which count the number of $u^+$ and $u^-$ harmonic variables respectively. The coefficients $f^{(i_1 \ldots i_n j_1 \ldots j_m)}$ are SU(2) tensors for which we can set

$$\Delta f^{(i_1 \ldots i_n j_1 \ldots j_m)} = (0,0). \tag{2.3}$$

The space of functions type Eq.(2.1) will be refereed hereafter to as $R_{[n+,m-]}$ and the total "ring" is simply given by

$$R = \bigoplus_{n,m \geq 1} \left( R_{[n+,m-]} + (R_{[n+,m-]})^* \right), \tag{2.4}$$

where $R^*$ stands for the dual space of R as we will discuss latter. We can also define the following product

$$\langle F^{[m+,n-]}(u), F_{[k+,l-]}(u) \rangle = \int du F^{[m+,n-]}(u) F_{[k+,l-]}(u)$$
$$= f^2 \frac{(-)^n m! n!}{(m+n+1)!} \delta_l^m \delta_k^n, \tag{2.5}$$

where we have used the $u^\pm$ integration rules [6,7]

$$\int du \, u^{[m+,n-]}(u) u_{[k+,l-]}(u) = \frac{(-)^n m! n!}{(m+n+1)!} \delta_l^m \delta_k^n \tag{2.6}$$

with

$$u^{[m+,n-]} = u^{+(i_1} \ldots u^{+i_m} u^{-i_{m+1}} \ldots u^{-i_{m+n})} \tag{2.7}$$

and

$$f^2 = f^{(i_1...i_n j_1...j_m)} f_{(i_1...i_n j_1...j_m)} \tag{2.8}$$

The introduced product Eq.(2.5), shows then that the spaces $R^{[n+,m-]}$ and $R_{[m+,n-]}$ are dual to each other. Using dimensional arguments, it is not difficult to see that the product $\langle\,,\,\rangle$ in Eq.(2.5) carries an harmonic weight $\Delta$ which we write for simplicity

$$\Delta(\langle\,,\,\rangle) \equiv -[(m+n)+,(m+n)-]. \tag{2.9}$$

Note by the way that the technical definition Eq(2.9) have nothing to do with the following feature

$$F_{[m+,n-]}F_{[k+,l-]} \neq F_{[(m+k)+,(n+l)-]} \tag{2.10}$$

which is a natural consequence of the standard reduction formulas [7]

$$u_i^+ u_{(j_1}^+...u_{j_n}^+ u_{k_1}^-...u_{k_m)}^- = u_{(i_1}^+ u_{j_1}^+...u_{k_m)}^- + \frac{m}{m+n+1}\varepsilon_{i(k_1} u_{j_1}^+...u_{j_n}^+ u_{k_2}^-...u_{k_m)}^- \tag{2.11}$$

$$u_i^- u_{(j_1}^+...u_{j_n}^+ u_{k_1}^-...u_{k_m)}^- = u_{(i}^- u_{j_1}^+...u_{k_m)}^- - \frac{n}{m+n+1}\varepsilon_{i(j_1} u_{j_2}^+...u_{j_n}^+ u_{k_1}^-...u_{k_m)}^- \tag{2.12}$$

Later on, we shall introduce a degrees pairing product ( , ) and a combined 'scalar' product $\langle\langle,\rangle\rangle$, built out of the harmonic weight product Eq.(2.5) and ( , ).

Furthermore, since the harmonic weight product get induces here, Eq.(2.4) can be written as

$$R = \bigoplus_{m,n\geq 1} \left[ R_{[m+,n-]} \oplus R^{[n+,m-]} \right] \tag{2.13}$$

## 3. The algebra of harmonic pseudo-differential operators

### 3.1 The harmonic space $\Xi_{[m+,n-]}^{(p,q)}$

To start, consider functions like

$$F^{[2n+,2m-]}(u) = f_{(i_1...i_{2n} j_1...j_{2m})} u^{+(i_1}...u^{+i_{2n}} u^{-j_{1}}...u^{-j_{2m})} \tag{3.1}$$

which will be denoted hereafter to as

$$F^{[2n+,2m-]}(u) = J^{[2n+,2m-]} \tag{3.2}$$

These kinds of functions will play the role of the analytic (conserved) fields, which are needed in the construction of Lax operators. A typical example of the analytic fields Eq(3.2) is given by $F^{[2+,(-2)-]}(u) = J^{[2+,(-2)-]}$ suspected to define the analogue of the KdV field of conformal spin 2, in the harmonic superspace language. The associated Lax operator (or harmonic KdV operator) is simply given by

$$L = D^{++2} + J^{[2+,(-2)-]} \tag{3.3}$$

with $D^{++} = u^+ \partial/\partial u^-$. This KdV like operator is suspected also to play an important role in the study of Liouville theory formulated in harmonic superspace language. A natural generalisation of the Lax operator Eq(3.3) is given by

$$d^{(p,q)}_{[a+, b-]} = \sum_{k=0}^{q-p} J^{[(a-q+k)+, (b+q-k)-]} D^{++(q-k)}, p \geq q \geq 0 \quad , \tag{3.4}$$

where the $J^{[(a-q+k)+, (b+q-k)-]}$'s are (1+q-p) analytic fields of harmonic weight $\Delta = ((a-q+k)+, (b+q-k)-)$ and where the p and q, with q ≥ p are integers that we suppose positive for the moment.

A particular example of the harmonic general Lax operator Eq(3.4) is given by setting a=-b=q as follows

$$d^{(p,q)}_{[q+, (-q)-]} = \sum_{k=0}^{q-p} J^{[(k)+, (-k)-]} D^{++(q-k)}, p \geq q \geq 0 \quad , \tag{3.5}$$

Using Eq(3.5) one can easily recover the form of the the harmonic KdV Lax operator, by setting just p=0 and q=2 namely

$$d^{(0,2)}_{[2+, (-2)-]} = J^{[0+, 0-]} D^{++2} + J^{[1+, 1-]} D^{++} + J^{[2+, (-2)-]}, \tag{3.6}$$

which coincides with the KdV like operator Eq(3.3) once the following choices are done

$$J^{[0+, 0-]} = 1 \text{ and } J^{[1+, (-1)-]} = 0. \tag{3.7}$$

On the other hand, the harmonic derivatives $D^{++}$ in Eq.(3.4) is considered in this analysis as an operator whose action on the analytic fields of the ring R is given by

$$D^{++} F = (D^{++} F) + \alpha F D^{++} \tag{3.8}$$

where $\alpha$ is some coefficient which we can explicitly determine. In order to derive explicit relations giving the action of the harmonic derivative $D^{++}$ and its higher power, on arbitrary analytic functions (Leibnitz rules) we proceed as follows. First we recall the following useful relations [15],

$$\frac{1}{D^{++}} u^+_{(i_1} ... u^+_{i_n} u^-_{j_1} ... u^-_{j_m)} = \frac{1}{m+1} u^+_{(i_1} ... u^+_{i_{n-1}} u^-_{i_n} u^-_{j_1} ... u^-_{j_m)} \text{ for } n \geq 1 \tag{3.9}$$

which gives the action of the operation $(D^{++})^{-1}$ for symmetrised u variables. As shown also in [15], the operation $(D^{++})^{-1}$ allows in general, additional terms of the form $(u^+)^{n-m}$ for n≥m because they are annihilated by $(D^{++})^{-1}$. Equation (3.9) is very important in the sense that it can leads, among others, to introduce pseudo-differential operators similarly to the standard Gelfand-Dickey formalism. This suspicion is motivated by the fact that one have always

$$D^{++} \frac{1}{D^{++}} F(u) = F(u) \tag{3.10}$$

for any function F on which $(D^{++})^{-1}$ can operate, and that

$$\frac{1}{D^{++}}D^{++}G(u) = G(u) \tag{3.11}$$

for any function G whose components all involve $u^-$. In order to do that, one need now to extend the definition Eq(3.9) to the operator form. Using the convention notation (2.7) we have presented in sect.(2), one can rewrite Eq(3.9) as

$$\frac{1}{D^{++}}u^{[n+,m-]} = \frac{1}{m+1}u^{[(n-1)+,\,(m+1)-]} \text{ for } n \geq 1 \tag{3.12}$$

which means creating one harmonic variable $u^-$ and annihilating a variable $u^+$ when acting on the symmetrised product of mixed harmonic variables $u^{[n+,m-]}$. Having in mind to write Leibnitz rules for the harmonic derivatives in the operator form, we use Eq(3.9), the properties Eqs(3.10-11) and the following ansatz

$$D^{++}u^{[n+,m-]} = (m+1)u^{[(n+1)+,\,(m-1)-]} + \alpha(m)u^{[n+,m-]}D^{++} \tag{3.13}$$

$$\frac{1}{D^{++}}u^{[n+,m-]} = \frac{1}{m+1}u^{[(n-1)+,\,(m+1)-]} + \beta(m)u^{[n+,m-]}\frac{1}{D^{++}} \tag{3.14}$$

where $\alpha(m)$, $\beta(m)$ are non vanishing coefficients depending only on the parameter index m. To determine these coefficients functions $\alpha(m)$ and $\beta(m)$ we require furthermore that

$$\frac{1}{D^{++}}D^{++}u^{[n+,m-]} = D^{++}\frac{1}{D^{++}}u^{[n+,m-]} \tag{3.15}$$

We find

$$\alpha\beta + \frac{m+1}{m} = \alpha + \frac{m+2}{m+1} \tag{3.16}$$

or simply

$$\alpha(1-\beta) = \frac{1}{m(m+1)} \tag{3.17}$$

Eq(3.17) shows among others that the function $\beta(m)$ can never take the value 1 as we can easily check from Eq(3.16). We learn also from this equation that the knowledge of one of the coefficient functions induces naturally the other. A simple choice consists for example in setting $\alpha = 1$ which gives $\beta = 1 - \frac{1}{m(m+1)}$. Next, we will not use such choice and we will work in the general case. On the other hand as shown in the ansatz Eqs(3.13-14), the harmonic derivatives $(D^{++})^{-1}$ and $D^{++}$ behave then as pseudo-differential operators with one as the inverse of the other by virtue of the constraint equation (3.15). They are shown also to behave formally like objects of harmonic weights

$$\Delta(D^{++})^{-1} = [(-1)+,\,(1)-] \tag{3.18}$$

and

$$\Delta D^{++} = [(1)+, (-1)-] \qquad (3.19)$$

Having extended the definition of the operation $(D^{++})^{-1}$ Eq(3.9) to the operator form Eq(3.14) and defining the action of its inverse $(D^{++})$ Eq(3.13), we are now in position to generalise the obtained result to higher power of these derivatives. Lengthy and straightforward algebraic manipulations lead to following important formulas

$$(D^{++})^p u^{[n+,m-]} = \sum_{k=0}^{p}(p-k)!\, c_{m+1}^{p-k} \sum_{m-p+k \leq j_1 \leq j_2 \leq \ldots \leq j_k \leq m} \alpha(j_1)\ldots\alpha(j_k) u^{[(n+p-k)+,(m-p+k)-]} (D^{++})^k \qquad (3.20)$$

$$(\frac{1}{D^{++}})^p u^{[n+,m-]} = \sum_{k=0}^{p}(p-k)!\, c_{m+1}^{m+p-k} \sum_{m \leq j_1 \leq j_2 \leq \ldots \leq j_k \leq m+p-k} \beta(j_1)\ldots\beta(j_k) u^{[(n-p-k)+,(m+p+k)-]} (\frac{1}{D^{++}})^k \qquad (3.21)$$

Some remarks are in order. Note first that when setting p=1, one recovers in a natural way the first leading relations of Eq.(3.20-21) namely Eqs(3.13-14) where the $\alpha$'s and $\beta$'s coefficients are given by Eq.(3.17). Note also that the Leibnitz rules equations (3.20-21) formulated in the harmonic superspace language present some originality compared to the standard Leibnitz rules formulated in the complex plane. Indeed, recall that the latter's are shown to incorporate in the case of negative powers of the derivatives $\partial_z = \partial/\partial z$ terms which go to infinity as shown in the following non local expression

$$\partial^{-k} f(z) = \sum_{l=0}^{\infty} (-)^l c_{l+k-1}^l f^{(l)}(z) \partial^{-l-k} \qquad . \qquad (3.22)$$

When trying to extend these standard formulas to the superspace harmonic language, we suspected to find relations for which the action of $(D^{++})^{-p}$ on arbitrary functions $F(u^{\pm})$ contains terms governed by the following expansion rule $\sum_{l=0}^{\infty}\ldots$ As shown in Eq(3.21), there is no terms at infinity and so there is no trivial way to go from standard relations to the harmonic case. The Leibnitz rules Eqs(3.20-21) give rise to local terms for both positive powers as well as for the negative powers of the harmonic derivatives. The harmonic case is then characterised by the local behaviour of its derivatives and contrary to Eq(3.22) no non local derivative is shown to appear.

Furthermore the harmonic Leibnitz rules Eqs(3.20-21) can be written for arbitrary functions belonging to the ring R as

$$(D^{++})^p F^{[n+,m-]} = \sum_{k=0}^{p} f_{(i_1\ldots i_n j_1\ldots j_m)} (p-k)!\, c_{m+1}^{p-k} \sum_{m-p+k \leq t_1 \leq t_2 \leq \ldots \leq t_k \leq m} \alpha(t_1)\ldots\alpha(t_k) u^{[(n+p-k)+,(m-p+k)-]} (D^{++})^k \qquad (3.23)$$

$$(\frac{1}{D^{++}})^p F^{[n+,m-]} = \sum_{k=0}^{p} f_{(i_1...i_n j_1...j_m)} (p-k)! C_{m+1}^{m+p-k} \sum_{m \leq t_1 \leq t_2 \leq ... \leq t_k \leq m+p-k} \beta(t_1) ... \beta(t_k) u^{[(n-p-k)+,(m+p+k)-]} (\frac{1}{D^{++}})^k \quad (3.24)$$

with $u^{[m+,n-]} = u^{+(i_1)}...u^{i_m} u^{-i_{m+1}}...u^{-i_{m+n}}$ as given in Eq.(2.7). To simplify much more these relations, let us introduce the following formal definition

$$F^{[(n+p-k)+,(m-p+k)-]} \equiv f_{(i_1...i_n j_1...j_m)} (p-k)! C_{m+1}^{p-k} \sum_{m-p+k \leq t_1 \leq t_2 \leq ... \leq t_k \leq m} \alpha(t_1)...\alpha(t_k) u^{[(n+p-k)+,(m-p+k)-]},$$

(3.25)

which gives

$$(D^{++})^p F^{[n+,m-]} \equiv \sum_{k=0}^{p} F^{[(n+p-k)+,(m-p+k)-]} (D^{++})^k, \quad (3.26)$$

$$(\frac{1}{D^{++}})^p F^{[n+,m-]} \equiv \sum_{k=0}^{p} F^{[(n-p-k)+,(m+p+k)-]} (\frac{1}{D^{++}})^k, \quad (3.27)$$

with

$$\Delta(D^{++})^p = (p+, p-). \quad (3.28)$$

Now, having defined how acts the harmonic derivatives $D^{++}$ and its inverse $(D^{++})^{-1}$ on the ring R of functions F, we are in position to build Lax operators extending the ones of Eq(3.4) to negative values of the degrees quantum numbers (p, q). The latter's are shown to have the following form

$$\delta_{[2q+,2m-]}^{(2p,2q)} = \sum_{k=0}^{q-p} J^{[2k+,2m-]} D^{++(q-k)}, \quad p \leq q < 0 \quad . \quad (3.29)$$

This configuration, which is a direct extension of Eq(3.4) is useful in the study of the algebraic structure of the space $\Xi_{[m+,n-]}^{(p,q)}$ and its generalisation to the space $\Xi^{(p,q)}$. Note also that, in analogy with the analysis developed in [14], one can use another representation basis of harmonic pseudo-differential operators namely, the Volterra-like representation. The latter is convenient in the derivation of the second Hamiltonian structure of generalised integrable hierarchies. Using Eqs(3.4,28) one easily see that operators with negative lowest degrees p and positive highest degrees q denoted $D_{[m+,n-]}^{(p,q)}$ can split as

$$D_{[m+,n-]}^{(p,q)} = \delta_{[m+,n-]}^{(p,q)} + d_{[m+,n-]}^{(p,q)} \quad . \quad (3.30)$$

More generally one have

$$D_{[m+,n-]}^{(p,q)} = \delta_{[m+,n-]}^{(p,k)} + d_{[m+,n-]}^{(k+1,q)}, \quad (3.31)$$

for any integers $p \leq k \leq q$.

## 3.2 The harmonic space $\Xi^{(p,q)}_{[0+,0-]}$

This space defines a limiting case for the spaces of harmonic pseudo-differential operators, since it supposes to correspond to differential Lax operators of vanishing harmonic weights Δ.

The Leibnitz product $*$ acts on the space $\Xi^{(p,q)}_{[m+,n-]}$, $q \geq p \geq 0$ as

$$* \; \Xi^{(p,q)}_{[m_1+,n_1-]} \times \Xi^{(p,q)}_{[m_2+,n_2-]} \to \Xi^{(p,2q)}_{[m+,n-]}, (m,n) = (m_1+m_2, n_1+n_2), \qquad (3.32)$$

upon performing the standard reduction formulas Eq(2.11). Eq(3.31) can be written in general as

$$* \; \Xi^{(p,q)}_{\Delta_1} \times \Xi^{(p,q)}_{\Delta_2} \to \Xi^{(p,2q)}_{\Delta_3}, \qquad (3.33)$$

where the total harmonic weight $\Delta_3$ is given by

$$\Delta_3 \equiv \Delta_1 + \Delta_2 \qquad . \qquad (3.34)$$

On the other hand one learn from Eqs(3.27) that the space $\Xi^{(p,q)}_{[m+,n-]}$ is not closed under the Leibnitz product $*$ not under the commutator of two harmonic differential operators. Imposing the closure, one gets strong constraints on the integers $m_i$, $n_i$, p and q, namely

$$[m_i+, n_i-]=[0+, 0-], \; 0 \leq p \leq q \leq 1 \quad , i=1,2; \qquad (3.35)$$

The particular space $\Xi^{(p,q)}_{[0+,0-]}$ satisfying the above constraints can exhibit then a Lie algebra structure provided that the Jacobi identity is fulfilled. Imposing to the Leibnitz product to be associative can ensure this. Note by the way the important fact that associativity in the space $\Xi^{(p,q)}_{[m+,n-]}$ is constrained to satisfy the standard reduction formulas Eqs.(2.11)., a property which is not, at first sight, easy to require as shown for example in the ring of harmonic functions.

## 3.3 The harmonic algebra $\Xi^{(p,q)}$

So far we have seen that the closure of the commutator of higher harmonic differential operators imposes constraints on the harmonic weights and on the degrees. The restriction on the harmonic weight can be overcome by using a larger set of operators referred hereafter to as $\Xi^{(p,q)}$ which is given by

$$\Xi^{(p,q)} = \bigoplus_{m,n \in N} \Xi^{(p,q)}_{[m+,n-]} \quad , p \leq q, \qquad (3.36)$$

The elements $D^{(p,q)}$ of this infinite dimensional space are differential operators with fixed degrees (p,q) but arbitrary harmonic weights. They read as

$$D^{(p,q)} = \sum_{m,n \in N} C(m,n) D^{(p,q)}_{[m+,n-]} \quad , p \leq q, \qquad (3.37)$$

where only a finite number of the c(m)'s are not vanishing. Setting for example p=q=0, we get the tensor algebra $R = \Xi_0^{(0,0)}$ discussed previously. As for the tensor spaces $\Xi_{[m+,n-]}^{(p,q)}$, the set $\Xi^{(p,q)}$ "can" exhibits a Lie algebra structure with respect to the commutator for $p \leq q \leq 1$.

### 3.4 The huge Lie algebra of harmonic operators $\Xi$

This is the algebra of harmonic differential operators of arbitrary harmonic weight and arbitrary degrees. It is obtained from the previous spaces by summing over all the allowed values of the degrees of the set $\Xi^{(p,q)}$. In some sense, it is the degree tensor algebra of $\Xi^{(p,q)}$ namely

$$\Xi = \bigoplus_{p \leq q} \Xi^{(p,q)} \quad , \tag{3.38}$$

Note that this infinite dimensional space is closed under the Leibnitz commutator without any constraint. Furthermore, a remarkable property of the space $\Xi$ is that it can splits into six infinite subalgebras given by $\Sigma_{\Delta_i}^{\pm}$, where $\Delta_i, i = -,0+$, refer respectively to harmonic weights which can be negative, zero or positive and where $\pm$ stands for the degrees quantum number which can be positive or negative. These 6=3x2 subalgebras are suspected to relate to each other's by some conjugation relations with respect to the degrees and harmonic weights. Before going to describe these conjugation features, let us now define the residue operation in the harmonic superspace language. In analogy with standard definitions of the residue operations in pseudo-differential operators, one propose the following definition

$$\text{Re}\, s \ (D^{++})^i = \delta_{i+1,0} \quad , \tag{3.39}$$

showing that the residue operation ($\text{Re}\, s \equiv \text{Re}\, s^{++}$) exhibits, in the harmonic superspace language, adapted to this analysis, two positive U(1) charges as explicitly shown in Eq(3.38). With this property, we see clearly that (Res.) carries a harmonic weight $\Delta = [(1)+, (-1)-]$ with respect to our convention notations. The principal consequence of this is the possibility to define an harmonic degrees pairing product ( , ) built out of the residue operation as we will show now. Let $d_{[a+,b-]}^{(p,q)}$ and $\delta_{[c+,d-]}^{(r,s)}$ defines respectively two arbitrary local differential and "pseudo"-differential operators given by

$$d_{[a+,b-]}^{(p,q)} = \sum_{i=p}^{q} X^{[(a-i)+,(b+i)-]} D^{++i} \tag{3.40}$$

$$\delta_{[c+,d-]}^{(r,s)} = \sum_{i=r}^{s} Y^{[(c-j)+,(j+d)-]} \left(\frac{1}{D^{++}}\right)^{-j} \tag{3.41}$$

The residue operation applied to the Leibnitz product of $d_{[a+,b-]}^{(p,q)}$ and $\delta_{[c+,d-]}^{(r,s)}$ is shown to be

$$\text{Re}\,s\left[d^{(p,q)}_{[a+,b-]} * \delta^{(r,s)}_{[c+,d-]}\right] = \sum_{i=p}^{q}\sum_{j=r}^{s}\sum_{k=0}^{i}\delta_{k+j,-1} X^{[(a-i)+,(b+i)-]}Y^{[(c-j+i-k)+,(d-i+j+k)-]} \quad (3.42)$$

where $\delta_{i,j}$ is the Kronecker index. Moreover Eq.(3.41) reads

$$\text{Re}\,s\left[d^{(p,q)}_{[a+,b-]} * \delta^{(r,s)}_{[c+,d-]}\right] = \sum_{i=p}^{q} X^{[(a-i)+,(b+i)-]}Y^{[(c+i+1)+,(d-i-1)-]} \quad (3.43)$$

which is a consistency algebraic expression if one knows that the harmonic weight of the l.h.s. of Eq.(3.42) is nothing but

$$\Delta(\text{Re}\,s[d^{(p,q)}_{[a+,b-]} * \delta^{(r,s)}_{[c+,d-]}]) = ((1+a+c)+,(-1+b+d)-). \quad (3.44)$$

coinciding exactly with that of the r.h.s. The previous algebraic computations show among others that the space $\Xi^{(p,q)}$ is dual to $\Xi^{(r,s)}$, once the following constraints on the degrees quantum numbers are satisfied

$$q+r+1=0 \text{ and } p+s+1=0. \quad (3.45)$$

Exploiting these results, one can then introduce the following harmonic degrees pairing product

$$(d^{(r,s)},\delta^{(p,q)}) = \delta_{1+r+q,0}\delta_{1+s+p,0} \text{Re}\,s\left[d^{(r,s)} * \delta^{(p,q)}\right] \quad , \quad (3.46)$$

which leads to decompose the huge space $\Xi$ as

$$\Xi = \Xi^- \oplus \Xi^+ \quad (3.47)$$

with

$$\Xi^+ = \bigoplus_{p\geq 0}\left[\bigoplus_{r\geq 0}\Xi^{(p,p+r)}\right] \quad , \quad (3.48)$$

$$\Xi^- = \bigoplus_{p\geq 0}\left[\bigoplus_{r\geq 0}\Xi^{(-1-p-r,-1-p)}\right] \quad . \quad (3.49)$$

The + and − down stairs indices carried by $\Xi^+$ and $\Xi^-$ refer respectively to the positive and negative degrees.

Having introduced the harmonic weight product Eq.(2.5) as well as the harmonic degree product Eq.(3.45), we are now in position to introduce a combined product $\langle\langle,\rangle\rangle$ given by

$$\left\langle\left\langle d^{(p,q)}_{[a+,b-]},\delta^{(r,s)|[b+,a-]}\right\rangle\right\rangle = \delta_{1+r+q,0}\delta_{1+s+p,0}\int du\, \text{Re}\,s\left[d^{(p,q)}_{[a+,b-]} * \delta^{(r,s)|[b+,a-]}\right] \quad (3.50)$$

from which we see easily that $\langle\langle,\rangle\rangle$, behaves as an object of harmonic weight $\Delta[\langle\langle,\rangle\rangle] = ((1)+,(-1)-)$.

The combined 'scalar' product relation Eq.(3.49) leads then to conclude that the spaces $\Xi^{(p,q)}_{[a+,b-]}$ are dual to $\Xi^{(-1-q,-1-p)[b+,a-]}$. The total number of these spaces is 6=3x2 depending on the values of the harmonic weight which can be positive, zero or negative and on the values (+, -) of the degrees quantum numbers. Among these spaces we note that the ones for which the quantum numbers ($\Delta$ and (p,q)) are positive with their dual having negative quantum numbers are very important in the construction of the Harmonic Gelfand Dickey Poisson bracket similarly as explicitly done in [14]. To close this section, we summarise here bellow some algebraic properties that we have used in the previous analysis.

|  | Harmonic Weight $\Delta$ | Degrees |
|---|---|---|
| $\Xi^{(p,q)}_{[b+,a-]}$ | (a+, b-) | (p , q) |
| $\Xi_{[0+,0-]}$ | (0+, 0-) | indefinite |
| $\Xi^{(p,q)}$ | indefinite | (p , q) |
| $D^{++p}$ | (p+,p-) | (p , p) |
| $\left(\dfrac{1}{D^{++}}\right)^p$ | ((-p)+,(p)-) | (-p, -p) |
| $\langle \Delta_1, \Delta_2 \rangle_{Eq.(2.5,9)}$ | $-\Delta_1 - \Delta_2$ | indefinite |
| Res. | (1+,1-) | indefinite |
| $\langle\langle , \rangle\rangle$ | (1+,1-) | indefinite |

## 4. Conclusion and discussion

Motivated by the believe that HS methods can lead to more explore the world of integrable physical problems, we have presented in this work the basic ingredients towards building in the harmonic superspace language the analogue of the GD algebra of pseudo-differential operators. This adaptation analysis necessitates at first sight to define the "ring" of analytic functions, which should describe the analogue of the standard conserved currents. In fact, using the results of [14], as well as the established harmonic superspace formulas Eqs(2.6,11,12), we was able to present the ring of harmonic functions, once some definitions and conventions notations are introduced. As a consequence of this construction, we have been able to describe also the space $\Xi$ of all harmonic Lax differential operators characterised by an harmonic weight $\Delta$ and two integers p and q describing respectively the lowest and the highest degrees. The present analysis is important also in the sense that it can leads to incorporates standard pseudo-differential calculus into the area of H.S. A lot of important and natural physical problems can follow this analysis. Let us mention some of them.

1. Use the present analysis to discuss the general integrability procedure of the following non linear differential equations on the sphere $S^2$

$$\partial^{++} q^+ - \partial^{++}\left[\frac{\partial V^{4+}}{\partial(\partial^{++}\bar{q}^+)}\right] + \frac{\partial V^{4+}}{\partial \bar{q}^+} = 0,$$

$$\partial^{++} \bar{q}^+ + \partial^{++}\left[\frac{\partial V^{4+}}{\partial(\partial^{++} q^+)}\right] - \frac{\partial V^{4+}}{\partial q^+} = 0,$$

where $q^+ = q^+(z,\bar{z},u^\pm)$ and its conjugates $\bar{q}^+ = \bar{q}^+(z,\bar{z},u^\pm)$ are complex fields defined on $CxS^2$ and where $V^{4+} = V^{4+}(q,u)$ is an interacting potential depending in general on $q^+, \bar{q}^+$ their derivatives and the $u^\pm$'s. Recall that the solvability of these equations is one the major problems of hyper-Kahler metrics building program [6,7,11,12,13]

2. Write the KdV equation and its higher non-linear generalisations in the H.S. language and discuss its Hamiltonian structures and the connection with 'conformal symmetry'.

3. Construct the G.D. Poisson bracket suspected to describe the second Hamiltonian structure of KdV integrable systems

4. Present a classification of 'harmonic' Lax operators and their various applications in physical non linear problems.

## Acknowledgements

The authors would like to thank the PARS program N 372/98 CNR who supported this research work. One of the authors M.B.S. would like to thank the good hospitality of the Abdus-Salm ICTP where a big part of this work has been done. He acknowledges also the help of the office of associate and federation schemes and the scientific help of the high-energy section. M.B.S. would like also to acknowledge the constant help of the Faculty of Science at Kenitra and the Department of Physics.